**OPEN**

# Measurement of 10 fs pulses across the entire Visible to Near-Infrared Spectral Range

Allan S. Johnson[1*], Emmanuel B. Amuah[1], Christian Brahms[2] & Simon Wall[1]

Tuneable ultrafast laser pulses are a powerful tool for measuring difficult-to-access degrees of freedom in materials science. In general these experiments require the ability to address resonances and excitations both above and below the bandgap of materials, and to probe their response at the timescale of the fastest non-trivial internal dynamics. This drives the need for ultrafast sources capable of delivering 10–15 fs duration pulses tuneable across the entire visible (VIS) and near infrared (NIR) range, 500– 3000 nm, as well as the characterization of these sources. Here we present a single frequency-resolved optical gating (FROG) system capable of self-referenced characterization of pulses with 10 fs duration across the entire VIS-NIR spectral range. Our system does not require auxiliary beams and only minor reconfiguration for different wavelengths. We demonstrate the system with measurements of pulses across the entire tuning range.

Ultrafast laser pulses have become an increasingly important tool in chemistry and materials science as their ability to separate couplings between relevant degrees of freedom in complex systems has been recognized[1–3]. The different degrees of freedom are generally energetically separate, and thus the capability to excite and measure the relevant physics depends critically upon the wavelength of the light used. The advent of commercial optical parametric amplifiers (OPAs) and more recently optical parametric chirped-pulse amplifiers (OPCPAs) has dramatically broadened the spectral range available in ultrafast laboratories away from the traditional 800 nm titanium sapphire or 1030 nm ytterbium lasers[4,5], allowing access to transitions above and below the bandgap for most complex materials. Accessing new spectral regions has led to dramatically new physics in many cases. In parallel, achievable pulse durations have continued to decrease, leading to routine generation of few-cycle pulses in the visible and near-infrared region, with duration down to one femtosecond or below possible[6,7]. Of particular interest are pulses of 10 fs duration, which are faster than nearly all vibrational[8] or magnetic[9] excitations as well as some electronic correlations[10], and thus capable of coherently exciting and capturing the fastest dynamics.

Despite the proliferation of sources and the need for experiments to pump and probe specific resonances across different spectral regions, pulse characterization of few-femtosecond pulses is usually done piecemeal, with pulse characterization devices typically constructed for a specific spectral window[11–14]. To the best of our knowledge only one system capable of measuring self-referenced pulses across the entire VIS-NIR has previously been demonstrated, but this system is not suitable for the measurement of ultrashort pulses due to the use of thick BBO crystals to upconvert the long-wavelength pulses under test to more convenient, shorter wavelengths[15]. For the high energy (μJ-mJ level) 10 fs duration, widely tuneable pulses increasingly common in ultrafast laboratories, a general solution is currently lacking. Here we present a device capable of measuring 10 fs full width half-maximum (FWHM) duration pulses across the entire VIS-NIR spectral region, providing such a general solution for high-power ultrafast materials science and spectroscopy laboratories.

For few-cycle pulses, spectral-phase interferometry for direct electric field reconstruction (SPIDER)[16], frequency-resolved optical gating (FROG)[13], and dispersion scan (d-scan)[17] have all been used in the VIS-NIR and have all proven to be excellent methods for pulse reconstruction. However, we believe that the FROG technique is particularly suited for measuring the broadly tuneable femtosecond pulses available in a growing number of labs for the following reasons. The dispersion control required by d-scan, particularly the need to tune through the point of maximum compression[18,19], makes it difficult to tune the wavelength away from those where sophisticated optical elements like chirped mirrors exist[20]. In contrast FROG and SPIDER have been used at a wide range of wavelengths. Other techniques from attosecond science can directly measure the electric field at the cost of significantly higher experimental complexity, in particular CEP stability[21–23].

[1]ICFO – The Institute of Photonics Sciences, The Barcelona Institute of Science and Technology, 08860, Castelldefels, Barcelona, Spain. [2]School of Engineering and Physical Sciences, Heriot-Watt University, Edinburgh, EH14 4AS, UK. *email: allan.johnson@icfo.eu

In SPIDER, narrowband and frequency-shifted auxiliary pulses are used to create upconverted replicas of the test pulses which then interfere, creating fringes which encode the spectral phase of the test pulse[16]. The auxiliary pulses can be chosen such that the upconverted pulses lie in a convenient frequency range, such as the absorption window of silicon (200–1100 nm) where high quality and cost effective CCD spectrometers are available. For instance mixing with 800 nm pulses means the entire range 500–3000 nm is mapped to the range between 300 nm and 650 nm. However, the use of auxiliary pulses may affect the portability and thus utility of SPIDER as it requires synchronizing multiple different colour pulses at different points where the device is used. This can be avoided if the auxiliary pulses are derived from the test pulse itself, but in this case the advantage of upconversion into the spectrometer sensitivity window is lost. In FROG the cross-correlation of the test-pulse and a gate is measured and frequency resolved, recording spectra as a function of delay. This time-frequency trace is then fed into an iterative algorithm to reconstruct the pulses[24]. In X-FROG the gate pulse is an independent pulse, in which case it can take advantage of the same upconverted spectral range discussed above for SPIDER, with similar drawbacks to portability and the additional drawback that the second pulse must be of comparable duration to the test-pulse. In most FROG configurations the gate is simply a copy of the test pulse modified by the non-linearity used. In the most common FROG configuration used, second-harmonic generation (SHG) FROG, the pulse and gate are identical and the spectrum of the fundamental is mapped to its second harmonic. Other commonly used configurations are third harmonic generation (THG), where the gate is the square of the test pulse, polarization gating, or self-diffraction[25]. Polarization gating and self-diffraction do not modify the wavelength of the signal from that of the test pulse, and thus are not appropriate for measurements spanning well over the range of any individual spectrometer. SHG maps the wavelengths 500–2200 nm onto the absorption of silicon (200–1100 nm) but wavelengths beyond this are lost. In contrast THG maps 600–3000 nm to the absorption of silicon, but shorter wavelengths cannot be captured. Each of these approaches has been used to measure ultrashort pulses, but a general solution has yet to be presented.

Here we present a single FROG device capable of measuring 10 fs full width half-maximum (FWHM) duration pulses across the entire VIS-NIR spectral region using a hybrid SHG-THG scheme. By switching between SHG and THG as the FROG non-linearity the full bandwidth required for 10 fs pulses is always mapped onto the wavelength range of a commercial silicon spectrometer. A scanning configuration allows both few-cycle and longer pulses to be measured without compromise, and the self-referenced nature of the measurements allow the device to be fully self-contained and portable. We demonstrate the FROG by measuring pulses across the full VIS-NIR, with few-cycle pulses characterized at some representative wavelengths in the visible and NIR. Our approach is general, and will allow many ultrafast laboratories to use a single device to characterize all available sources in a direct manner at their point of use.

## Setup

We opt for a non-collinear scanning FROG configuration, where the pulse is split into two replicas. A translation stage is used to control the relative delay of the replicas, which are then focused into the non-linear medium at a small relative angle. It has been previously shown that because the temporal blurring from the crossing angle adds in quadrature to the retrieved pulse duration in FROG, blurring of even a few femtosecond introduces only a few-hundred attosecond error to the retrieved FROG trace, and this effect is negligible down to the 10 fs level for an appropriately chosen crossing angle[26]. We use division of wavefront to split the input beam into two copies for self-referenced measurements, in this case by a simple d-cut silver mirror. One beam passes through a delay arm capable of 100 nm steps and the two are re-combined co-linearly with a slight spacing (approximately 2 mm) by another d-cut mirror. The two beams then reflect off a 15° off-axis parabolic mirror, 380 mm focal length, and into the generation medium where they overlap at the focus. The generated signal and fundamental then propagate to a 15 cm focal length spherical aluminium mirror 30 cm away which re-images the crossing point onto the entrance slit of a CCD spectrometer covering the range 200–1050 nm with 0.1 nm resolution (Avantes Minispec). The combined large bandwidth and high spectral resolution is essential for allowing the measurement of both ultrashort broadband pulses and longer, narrowband pulses, like those used in THz generation. Aluminum mirrors are used after the generation medium to ensure good reflectivity in the UV for second-harmonic generation of visible pulses. An adjustable slit in front of the mirror is used to select the signal beam from the fundamentals and other non-linear signals[27]; the complete system is shown in Fig. 1.

Only two simple adjustments are required to switch from measuring pulses across the entire visible range and much of the shorter wavelengths NIR to measuring the longer wavelength NIR. The first is to change the generation medium, moving from SHG for shorter wavelengths to THG for the longer wavelengths. The second is a minor adjustment to the adjustable slit to compensate the different direction of the emitted signal. Because the medium is used in transmission and the crossing point is re-imaged to the spectrometer no other alignment is required; by using magnetic bases to swap the nonlinear media we were able to switch from one configuration to the other in under a minute.

We first consider SHG for the shorter wavelengths. There is extensive and illuminating literature on the use of SHG for the measurement of pulses down to single-cycle in duration[13,27]; SHG benefits from high signal levels, no competing non-linearities of the same order, and extremely robust reconstruction algorithms. The significant challenge is obtaining sufficient phasematching bandwidth to accurately upconvert the full bandwidth of a 10 fs pulse across the entire VIS-NIR region. Simultaneously fulfilling this requirement across this spectral range is unfortunately not possible for any known material; we instead use a thin cut (5 micron) crystal of BBO and scan the phase-matching angle for different wavelengths. This thickness proved to provide sufficient signal even to measure the nanojoule pulses from an oscillator. By orienting the extraordinary axis in the plane perpendicular to the incident polarization the phasematching condition can be easily tuned by rotating the crystal. This results in a slight change in effective crystal thickness and effective angle through the crystal due to refraction, but because the incident radiation is polarized purely along an axis with normal refractive index this factor can be easily included. Refraction additionally translates the beam across the spectrometer entrance slit as the crystal is rotated. This is reduced so as to be negligible by cutting the crystal at an intermediate angle - in our case a cut at 40 degs ensured that no wavelength required tuning more than 20 degrees off normal. The phasematching bandwidth for this configuration is shown in Fig. 1b, with the calculation including the deflection angle and change in effective thickness for the fundamental. The cut angle required for phasematching a 10 fs pulse varies much more rapidly in the visible than in the NIR, where a single angle can phasematch almost the entire range. Note that no angle dithering[28] is required to obtain sufficient bandwidth even at the shortest wavelengths of 500 nm – the angle needs only be set depending on the input wavelength.

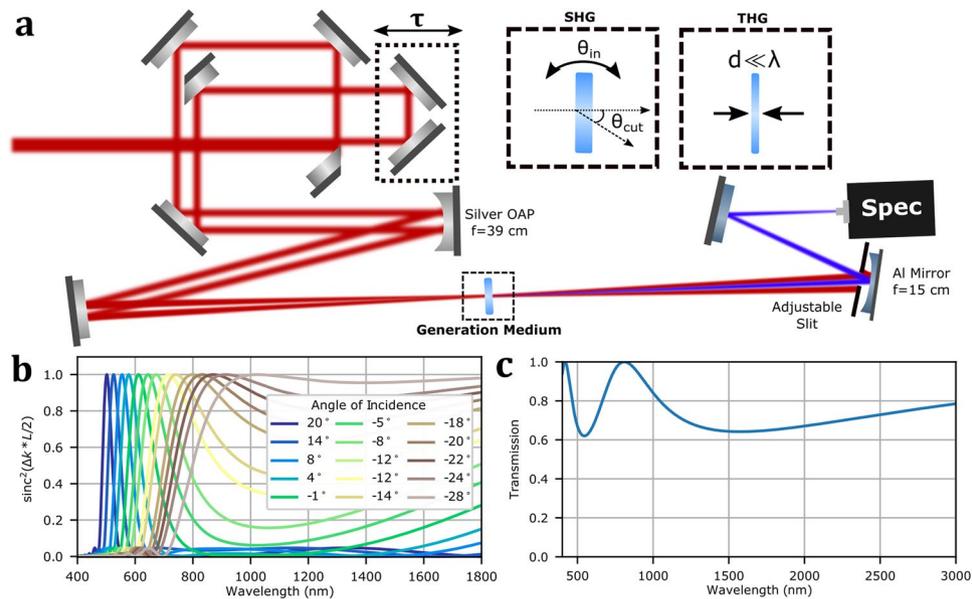

**Figure 1.** (**a**) Diagram of hybrid SHG/THG scanning FROG for full VIS-NIR coverage. The input beam is geometrically split and a delay introduced in one arm. The beams are focused by a silver off-axis parabolic mirror into the generation medium. The generated signal is separated from the fundamentals and secondary signal (in case of THG) by an adjustable slit. The signal is then re-imaged on the entrance slit of a spectrometer. Insets: BBO for SHG and thin film of $Si_3N_4$ for THG. (**b**) Phasematching between f and 2f frequencies in 5 micron thick BBO as a function of fundamental frequency f for various incidence angles, showing the broad tuneability and large supported bandwidths. Here 0 degrees is the beam entering at normal incidence, with an angle to the extraordinary axis of 40 degrees. (**c**) Calculated transmission through 200 nm thick film of $Si_3N_4$, showing minor etalon effects which should be included in the response function for the FROG reconstruction.

For longer NIR wavelengths (those with wavelength components beyond 2200 nm) we use THG FROG. THG has the advantage of large upconversion bandwidth, with THG from material surfaces having previously been used for FROG measurements[29]. The THG is strongly confined to the surfaces because of group velocity and phase mismatch, and thus there is minimal distortion of the harmonic signal from macroscopic effects[30] provided the crystal is sufficiently thick that the contribution from the back surface is suppressed by beam walkoff[31]. We have previously used a thick piece of fused silica for THG FROG measurements of long (100 fs) pulses in this manner. However, this approach was found to fail for shorter pulses because while the THG is confined to the first few wavelengths of the material, self- and cross-phase modulation (SPM, XPM) of the fundamental take place throughout the sample, allowing significant spectral bandwidth to extend into the frequency window of the FROG trace and, through the XPM, scatter into the direction of the third-harmonic emission. Here we use a 200 nm thick free-standing thin-film of silicon nitride ($Si_3N_4$) as the third harmonic medium instead. Free-standing films of multi-micron thickness[32] or nanolayers deposited onto substrates[33] have been previously used for THG FROG, but to the best of our knowledge this is the first time a sub-wavelength free-standing film has been used to remove phasematching effects and supress competing non-linearities. $Si_3N_4$ is an ideal material due to its robust free-standing nature, high damage threshold, and lack of significant resonances in the VIS-NIR. There are only minor etalon effects from reflections inside the film itself, as shown in Fig. 1c, and these can be included in trace calibration. Because THG is naturally restricted to only a small longitudinal region, the overall drop in efficiency relative to a thick sample is significantly lower than may be expected, and energies of a few-tens of microjoules are sufficient to generate good signal levels.

## Results

To demonstrate the system we measured a wide range of pulses generated from an amplified titanium-sapphire laser system (Coherent Astrella) and visible and infrared optical parametric amplifiers (OPAs) (Light Conversion). The Ti:saph system provided amplified narrowband pulses of sub-100 fs duration and broadband oscillator pulses of less than 30 fs transform limited duration at 800 nm, while the non-collinear visible OPA provided pulses of down to 10 fs in duration tuneable from 500 nm to 1100 nm. The infrared OPA provided longer pulses tuneable from 1140 nm to 2700 nm; we used subsequent hollow-core fibre pulse compression to reduce the duration of pulses in the range 1600–2100 nm down to 10–20 fs[11,14]. For each measurement the input energy and acquisition exposure time was adjusted to nearly saturate at zero relative delay to give the highest dynamic range. 128 sampling points were taken extending across the range where the signal was above the noise and into the zeros on either side. This leads to a varying step size for different pulse measurements. Measurements took between one minute and 15 minutes, depending on pulse energy. The noise floor was generally around 0.3% of peak for our spectrometer dynamic range, rising to 1% for low energy pulses, and was thresholded prior to reconstruction. Stability of the reconstructions to resampling was tested for each trace.

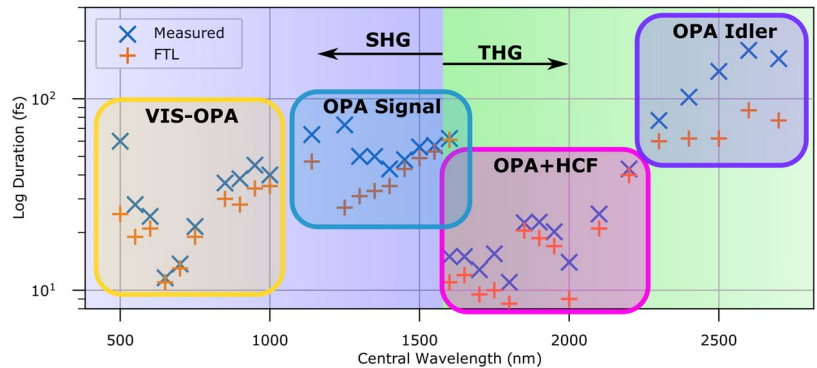

**Figure 2.** Pulse durations measured across the entire VIS-NIR wavelength range. Blue exes are best achieved pulse duration while orange crosses are the transform limited FWHM. The FROG error is below 0.003 for all SHG measurements and 0.005 for all THG measurements.

Figure 2 shows the pulse durations measured across the entire VIS-NIR range. Measurements were repeatable over the course of a week. The reconstructions are uniformly good, in particular with FROG errors less than 0.005 for all of the more challenging THG measurements. Pulses of below 15 fs in duration are measured between 650–700 nm direct from the VIS-OPA and between 1600–2000 nm when compressed with the HCF; elsewhere we are not limited by the measurement but rather by our ability to generate 10 fs pulses. Large variations are observed when switching between different pulse generation techniques – the uncompressed signal of the OPA at 1600 nm is approximately 6 times longer than the compressed idler at the same wavelength. FROG traces were corrected for the spectral response of the system taking into account the spectrometer response, non-linear coefficients $\chi$, and phasematching bandwidths (SHG only)[27]. The response of the spectrometer and the routing optics after the nonlinear media was calibrated by imaging a thermal source to the crystal plane and measuring the spectrum, then correcting the spectrum to match the known thermal emission curve. While the thermal source is unpolarised, the spectrometer will likely have a polarization dependent response, leading to a possible source of systematic error in our measurements. As the THG and SHG measurements have different polarizations, an unpolarised source gives the best compromise response curve. Given a sufficiently broadband polarizer it should be possible to isolate one polarization component of the thermal source and calibrate the two input polarization response curves separately. The second order non-linearity of BBO was estimated using Miller's rule[34] and the third-order non-linearity of $Si_3N_4$ was estimated from its linear properties as described by Boling et al.[35]. Failure to include this correction factor resulted in poor agreement between measured and reconstructed traces, along with a failure to accurately predict the effect of additionally added material dispersion.

All measurements below 1600 nm were performed with SHG FROG. While we measured near 10 fs pulses at 650 nm and 700 nm, our OPA is known to provide 10 fs pulses in this wavelength range only. We were thus unable to verify the capability to measure 10 fs pulses down to 500 nm central wavelength, but the dispersion curves of BBO are sufficiently well known that there is no reason to believe that the FROG would not function as designed[36]. Figure 3 shows the FROG trace of a 12 fs pulse centered at 700 nm plotted on a log-scale; the quality of the reconstruction, performed with the principal components generalized projections algorithm (PCGPA)[37] is excellent, with a FROG error of below 0.005. As an independent quality check, a comparison of the reconstructed spectrum to one independently measured by opening the signal selecting slits is also shown. We briefly note that the use of very thin BBO crystal to obtain large phasematching bandwidths is often criticized on the grounds that this leads to prohibitively low signal levels. We found however that 5 microns was sufficient to perform a high quality measurement of 5 nJ pulses from an oscillator at 80 MHz repetition rate, even when using a much longer focal length than is common in measurements of oscillator pulses.

For wavelengths between 1600 nm and 2100 nm we coupled the output of the OPA into a 50 cm long, 250 micron inner diameter hollow capillary filled with 2.8 bar of Ar to spectrally broaden the pulses. The broadened supercontinuua were compressed using a mixture of a transmissive longpass filter (FELH1100, Thorlabs), fused silica glass plates, and a pair of fused silica wedges, before being delivered to the FROG. Generally much more glass was required to compress short wavelengths, in keeping with previous measurements of this type of system[14]. The second row of Fig. 3 shows a typical THG-FROG trace taken at 1800 nm with 5.3 mm of fused silica and the longpass filter for post-compression, along with the reconstructed trace and pulse. A FROG error of 0.003 was obtained using an open source FROG software (Femtosoft) after resampling to a 512×512 grid and applying a hot-pixel filter. There was no appreciable change in performance of the FROG across the full range of the fibre output, nor for wavelengths between 2300 nm and 2700 nm in which we had to bypass the fibre and measure the uncompressed idler of the OPA. A comparison to an independently measured fundamental spectrum also shows good agreement; this spectrum, in contrast to the SHG case, cannot be measured *in-situ* and a separate measurement is made prior to the FROG.

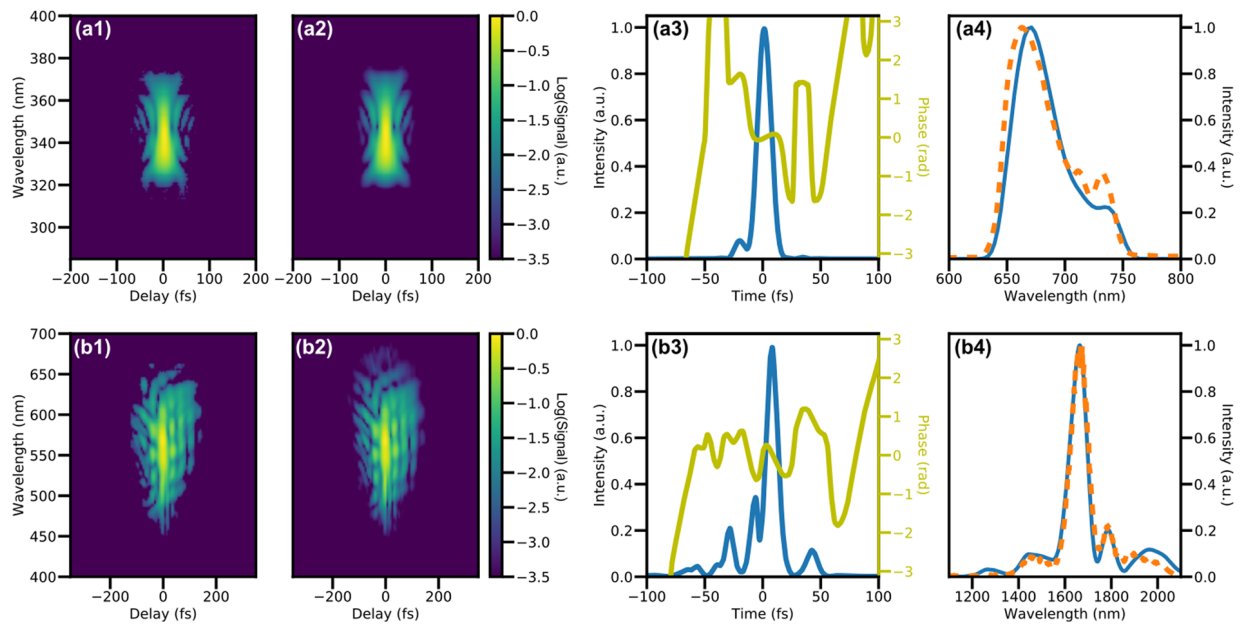

**Figure 3.** Experimental FROG traces and their reconstructions: Second harmonic FROG traces of a 12 fs pulse at 700 nm (top row, **a**) and THG FROG of an 11 fs pulse at 1800 nm (bottom, **b**). From left to right: experimental FROG trace (log10 scale) (1), reconstructed FROG trace (log10 scale) (2), retrieved pulse (3), and retrieved spectrum (blue) with independently measured spectrum (orange) (4).

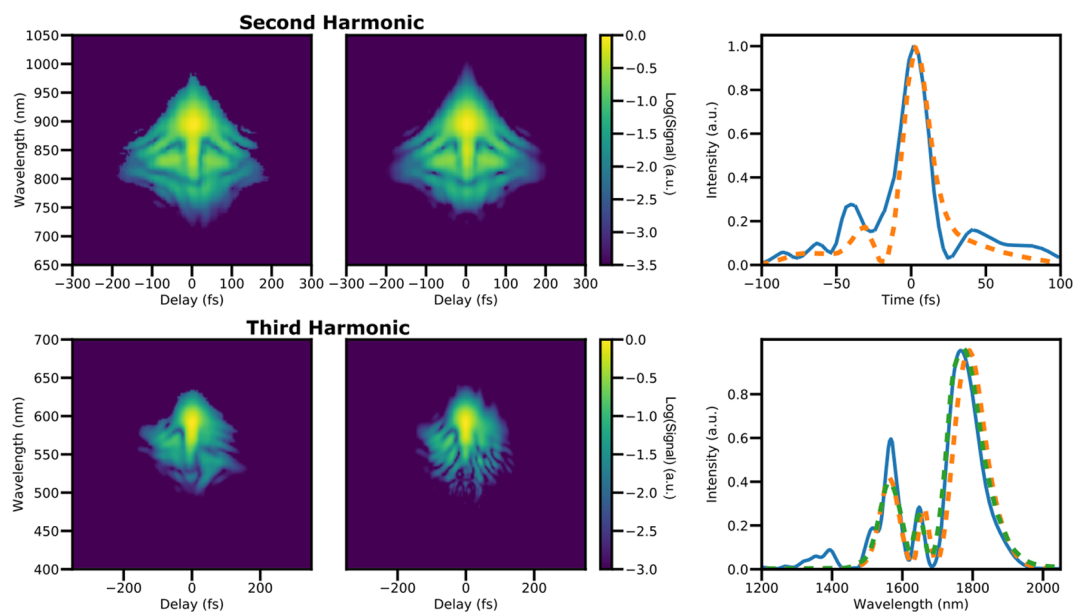

**Figure 4.** Comparison of SHG and THG FROG measurements on the same pulse, taken sequentially. Top Left and Top Center: experimental and reconstructed SHG FROG traces. Bottom left and center: experiment and reconstructed THG traces. Top right: retrieved pulses from SHG (orange) and THG (blue). Bottom right: retrieved spectrum from SHG (orange) and THG (blue) traces, as well as independently measured fundamental spectra (green).

We found that while THG-FROG performs well for simple, near Gaussian pulses or near transform limited pulses as found previously in the literature[29,32,38,39], all tested reconstruction algorithms struggled with broadband, chirped pulses such as those typically encountered from a HCF. Those algorithms tested include all those detailed by DeLong and Trebino[40] as implemented in the Femtosoft FROG software, as well as the PCGPA[37]. Because these chirped pulses have lower peak fields than the compressed pulses these problems cannot be attributed to non-linear effects distorting the test pulses or contaminating the FROG traces; for this reason, we attribute the challenge to the nature of the THG-FROG traces themselves. In practice this was not overly limiting, as a near transform limited pulse could be found by adding dispersive elements feeding back on the bandwidth of the upconverted signal at temporal overlap, then performing a full FROG trace to measure the pulse. Using this approach we were able to achieve convergence at the shortest pulse duration for each wavelength tested. More advanced reconstruction algorithms may alleviate this constraint[41,42], but to the best of our knowledge, the case of complex non-collinear THG-FROG traces has not been examined in detail.

Figure 4 shows reconstructions of the same input pulse using the SHG and THG configurations, along with a comparison of the retrieved fundamental spectra along with an independently measured one. The agreement is good between all methods. Switching between the SHG configuration and THG configuration required less than 1 min, demonstrating the ease of use of our device. Multiple alternating switches were made, showing no degradation in the quality of acquired traces, demonstrating the robustness of the scheme.

## Conclusions

We have demonstrated a single FROG device capable of characterizing 10 fs pulses centered at wavelengths across the entire VIS-NIR spectral range (500–3000 nm) using second and third harmonic generation. Moving from SHG to THG, used for short and long wavelength pulses, respectively, is done simply by switching the transmission generation medium and adjusting a slit used to select the signal. The system is entirely self-referenced and thus portable, capable of characterizing pulses at a variety of positions in the various experimental setups without reconstructing delay-lines for auxiliary pulses. We proved the utility by measuring pulses across the entire tuning range, with near 10 fs pulses achieved at various wavelengths in the visible and near-infrared. Our approach will be useful for ultrafast laboratories which increasingly use sources at a wide range of wavelengths but do not wish to construct a series of pulse-characterization devices.

## Data availability

The datasets generated during and/or analyzed during the current study are available from the corresponding author on reasonable request.

### Acknowledgements
This project has received funding from the European Research Council (ERC) under the European Union's Horizon 2020 research and innovation programme (Grant Agreement No. 758461) and was supported by Spanish MINECO (Severo Ochoa grant SEV-2015-0522, SEV2015-0496) as well as Fundació Privada Cellex, and CERCA Programme/Generalitat de Catalunya. This project has received funding from the European Union's Horizon 2020 research and innovation programme under the Marie Skłodowska-Curie grant agreement No. 754510. This work was funded by the European Research Council (ERC) under the European Union's Horizon 2020 research and innovation program (grant agreement No. 679649).

### Author contributions
A.S.J. and S.W. conceived the project. A.S.J. and E.A. performed the experiments. A.S.J. and C.B. processed and analyzed the FROG traces. All authors contributed to the manuscript.

### Competing interests
The authors declare no competing interests.

### Additional information
**Correspondence** and requests for materials should be addressed to A.S.J.

**Reprints and permissions information** is available at www.nature.com/reprints.

**Publisher's note** Springer Nature remains neutral with regard to jurisdictional claims in published maps and institutional affiliations.